# Unlocking CCUS-Ready Coal Power Investments: A Spatial Real Options Approach

Yun-Long Zhang[1,2,3,4], Jia-Ning Kang[1,2,3,5 *], Lan-Cui Liu[6], Yi-Zhuo Ji[1,2,3,5], Song Peng[1,2,3,5], Yi-Ming Wei[1,2,3,5 *]

**Abstract:** New coal-fired capacity may still be needed in some economies to support energy security, yet tightening carbon constraints increasingly threaten its long-term investment viability. Designing new plants to be Carbon Capture, Utilization and Storage (CCUS)-ready is therefore critical, but existing investment studies largely neglect the spatial factors that shape CCUS investment viability. In this study, we propose a Spatial Real Options (SRO) framework that couples national-scale siting screening with site-level real-options valuation. A national inventory of technically feasible sites across China (n = 194,027) is linked to a real-options model that incorporates temporal uncertainty together with province-specific conditions, $CO_2$ transport distance, and alternative policy instruments. Results show that the nationwide investable window for conventional coal closes by the early-to-mid 2040s. CCUS integration materially reshapes project economics, but outcomes differ substantially across storage options and transport distances: oil-reservoir storage remains economically attractive at short distances, whereas saline-aquifer storage requires operational incentives to become competitive. Generation-hour compensation substantially outperforms investment-cost subsidies, indicating that operating costs, rather than upfront capital, constitute the binding constraint. The resulting spatiotemporal maps provide a practical decision-support tool for identifying where and when CCUS-ready coal investments are most attractive under progressive decarbonization.

**Keywords:** Coal-fired power; CCUS-ready; Real options; Spatial analysis; Investment timing; Spatial Real Options (SRO)

[1] Center for Energy and Environmental Policy Research, Beijing Institute of Technology, Beijing 100081, China. [2] Beijing Lab for System Engineering of Carbon Neutrality, Beijing Municipal Education Commission, Beijing 100081, China. [3] NSFC Basic Science Center for Energy and Climate Change, Beijing 100081, China. [4] Division of Physical Resource Theory, Department of Environmental and Energy Sciences, Chalmers University of Technology, 412 96, Göteborg, Sweden. [5] School of Management, Beijing Institute of Technology, Beijing 100081, China. [6] School of National Safety and Emergency Management, Beijing Normal University, Beijing 100875, China.

*Corresponding authors: wei@bit.edu.cn (Y.-M. Wei), kangjianing@bit.edu.cn (J.-N. Kang)

## 1. Introduction

Global coal power additions have declined for a tenth consecutive year as numerous economies pledge to phase out coal. Nevertheless, residual demand for new coal capacity persists in regions where rapid electricity demand growth and energy-security concerns continue to require reliable and dispatchable generation. In 2024, eight countries proposed new coal-fired power projects. While China experienced a slowdown in new proposals, India reached a record high. China nonetheless continued to register historically high levels of construction starts and project approvals. These developments reflect (i) persistent concerns over energy security following the COVID-19 pandemic, extreme weather events (Jiang et al., 2023) and geopolitical conflicts such as the Russia–Ukraine war (Global Energy Monitor et al., 2024), and conflicts in the Middle East (Al-Sarihi, 2026; Block et al., 2025); (ii) the growing need for grid-stability and flexibility as renewable generation expands, often far from major load centers (Ullah et al., 2024); and (iii) the replacement of aging and inefficient generating units.

For economies pursuing stringent climate targets, however, security-driven coal investments face a structural dilemma. Under China's carbon peaking and carbon neutrality goals, declining utilization hours can undermine profitability for new units and increase the risk of asset stranding (Chen and Lin, 2024). If new coal capacity remains necessary to ensure system reliability and energy security, investment decisions must therefore consider both near-term profitability and long-term economic viability under progressively tightening carbon constraints.

Among available mitigation options, carbon capture, utilization, and storage (CCUS) is one of the few pathways capable of delivering deep emissions reductions while preserving dispatchable capacity (Zhang et al., 2024, 2023). As a result, increasing attention has been directed toward CCUS-ready coal power, in which new plants are designed to accommodate future carbon capture deployment when economic and policy conditions become favorable (Bukar and Asif, 2024). The viability of such investments depends not only on the profitability of the initial plant investment but also on whether future CCUS deployment becomes economically attractive under evolving carbon prices, fuel costs, operating conditions, technological progress, and policy support.

The investability of CCUS-ready coal power is inherently location-dependent. Technical feasibility determines where new projects can be developed through constraints such as land availability, protected areas, terrain, seismic risk, and proximity to electricity demand (Rehman et al., 2019; Uyan, 2017; Wang et al., 2018; Xu et al., 2020). Even among technically feasible sites, investment performance can differ substantially because of regional variations in fuel prices, electricity tariffs, plant utilization, and the geographic relationship between emission sources and suitable $CO_2$ storage resources(Fan et al., 2021, 2020). In particular, source–sink distance directly affects $CO_2$ transport costs, while storage options differ in both costs and revenue potential (Wang et al., 2020; Wei et al., 2022a; Wu et al., 2022). Consequently, projects with otherwise similar technical characteristics may exhibit markedly different investment values and optimal investment timings simply because of their location.

Existing studies have examined these issues from two complementary perspectives. Spatial analyses have investigated plant siting (Assen et al., 2016), source–sink matching (Wei et al., 2022a), $CO_2$ transport infrastructure (Sun and Chen, 2017; Wei et al., 2022a), and geological storage availability (Keating et al., 2011), identifying where CCUS deployment is technically or economically favorable. In parallel, real options (RO) models have been widely applied to energy investments to determine optimal investment timing under uncertain carbon prices, fuel costs, electricity prices, technology costs, and policy conditions (Fan et al., 2023; Tan et al., 2023; Wang and Zhang, 2018; Wu et al., 2013; Zhou et al., 2021). However, these two research streams remain largely separate. Spatial studies typically evaluate deployment potential without considering dynamic investment decisions, whereas RO analyses generally focus on representative projects or aggregated regional cases without explicitly incorporating site-specific spatial information into project valuation. Consequently, how location-dependent factors influence investment timing, investment value, and investment likelihood at the national scale remains insufficiently understood.

To address this gap, we develop a Spatial Real Options (SRO) framework to evaluate the spatiotemporal investability of CCUS-ready coal power in China. The framework combines national-scale spatial screening with site-specific real-options valuation by incorporating location-dependent characteristics into project cash flows under uncertainty. Technical feasibility is first assessed using a nationwide 5 km × 5 km spatial screening considering environmental, geophysical, land-use, and infrastructure constraints. For each feasible location, province-specific economic conditions and site-specific source–sink relationships are incorporated into an RO model that considers uncertainties in carbon prices, fuel prices, annual operating hours, technological progress, and alternative policy mechanisms. Applied to 194,027 technically feasible locations across China, the framework evaluates investability through three complementary indicators—optimal investment timing, investment probability, and expected project value—and reveals how spatial heterogeneity shapes investment opportunities under different storage and policy scenarios.

This study makes three principal contributions. **Methodologically,** it develops an integrated SRO framework that links national-scale spatial screening with site-specific investment valuation by allowing location-dependent characteristics to directly influence investment decisions under uncertainty. **Empirically,** by applying the framework to 194,027 feasible locations, it provides a national-scale assessment of the spatiotemporal investability of CCUS-ready coal power, revealing geographically differentiated investment windows and deployment opportunities that cannot be captured by representative-project analyses. **From a policy perspective,** the resulting investability maps help identify favorable investment locations and evaluate how different policy support mechanisms influence the economic attractiveness of CCUS-ready coal power, providing useful evidence for reducing stranded-asset risks while supporting energy security during China's low-carbon transition.

## 2. Methodology

### 2.1 Framework overview

We develop a Spatial Real Options (SRO) framework to evaluate the spatiotemporal investability of CCUS-ready coal power by combining national-scale spatial screening with site-specific real-options valuation under uncertainty (Fig. 1). The framework first identifies technically feasible locations for new CCUS-ready coal-fired power plants through spatial screening and then evaluates the investment timing and value of each feasible site under alternative CCUS storage and policy scenarios.

Spatial screening applies established siting constraints, including protected-area avoidance, topographic slope limits, land-use exclusions, seismic-risk thresholds, and proximity to load centers. For each feasible site, location-dependent characteristics are incorporated into the RO model through province-specific economic and operational conditions and site-specific source–sink relationships that determine $CO_2$ transport distance and associated costs, while temporal uncertainties and policy conditions are represented in the investment valuation. Applying the framework to all technically feasible sites (n = 194,027) produces spatially explicit estimates of optimal investment timing, investment probability, and expected project value, enabling national-scale assessment of the spatiotemporal investability of CCUS-ready coal power.

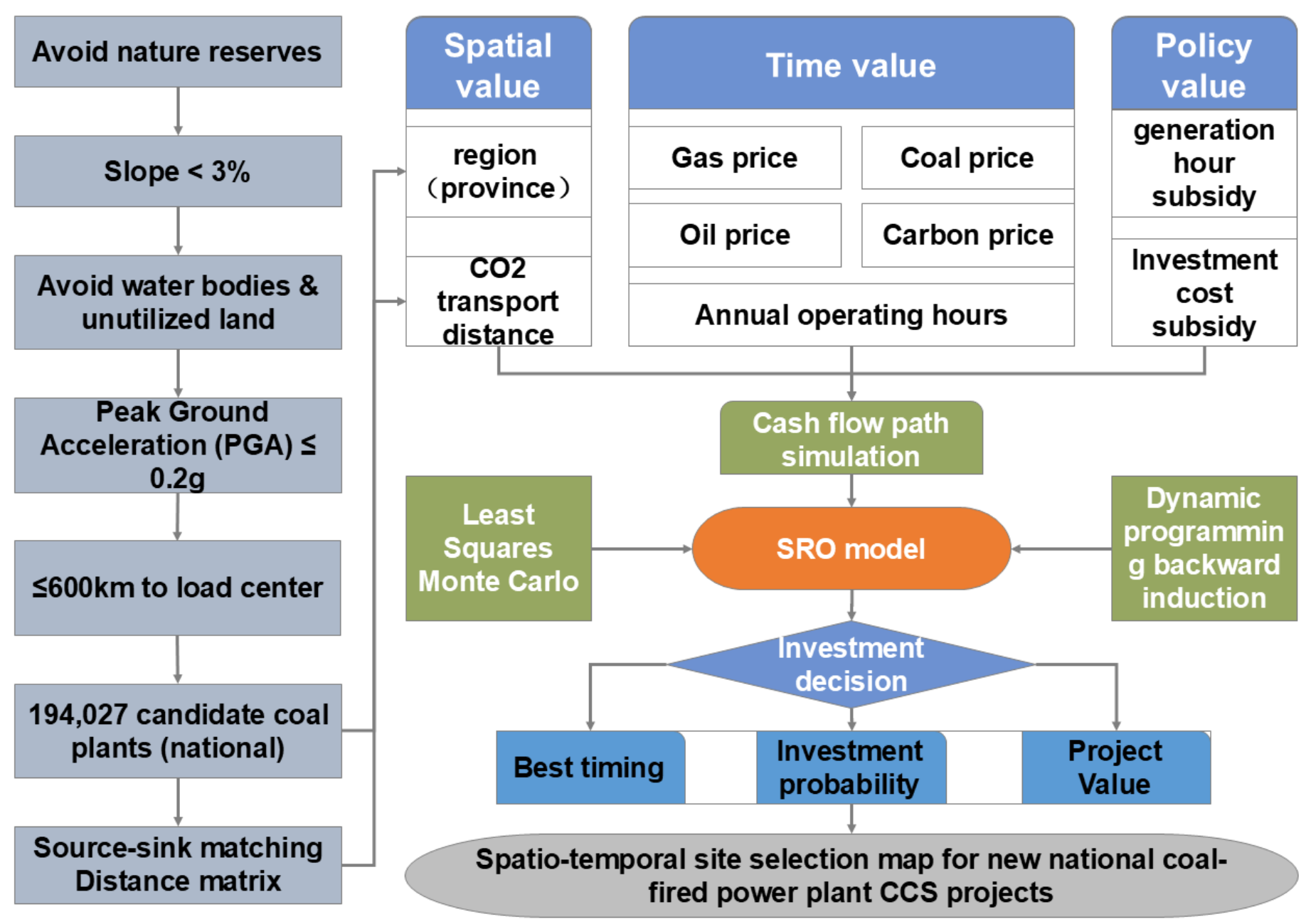

Fig. 1 Spatial Real Options (SRO) framework for evaluating the spatiotemporal investability of CCUS-ready coal power. The workflow combines national-scale spatial screening with site-specific real-options valuation under spatial heterogeneity and temporal uncertainty to generate national maps of optimal investment timing, investment probability, and expected project value.

## 2.2 Spatial screening of technically feasible sites

The first step of the SRO framework is to identify technically feasible locations for new CCUS-ready coal-fired power plants through national-scale spatial screening. A 5 km × 5 km grid was generated in ArcGIS, with a 1,000 MW ultra-supercritical coal-fired unit adopted as the reference plant configuration. Following the *Land Use Indicators for Construction of Electric Power Projects* issued by the relevant Chinese authorities, five siting criteria were applied, covering protected areas, topographic slope, land use, seismic risk, and proximity to load centers (Fig. 2).

Grid cells failing any criterion were excluded, leaving **194,027** technically feasible sites. The screening results were validated by comparing the identified feasible sites with the locations of operational, under-construction, and proposed coal-fired power plants. The resulting match rates were **83.5%** for operational plants and **88.7%** for proposed plants. The slightly lower match rate for operational plants likely reflects land-use changes since their commissioning relative to the ~2020 baseline of the spatial datasets, supporting the validity of the screening procedure.

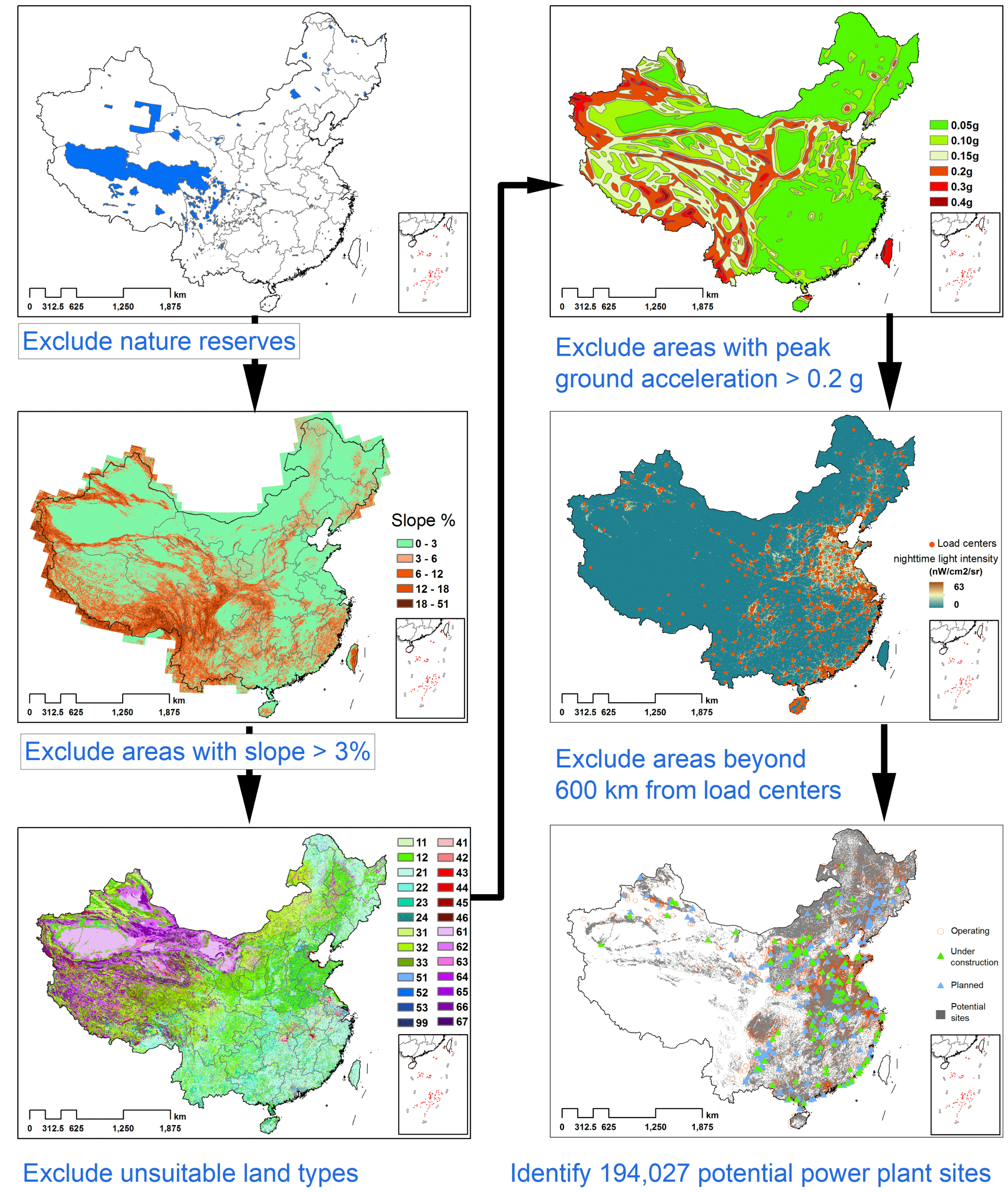

Fig. 2 National-scale spatial screening of technically feasible sites for new CCUS-ready coal-fired power plants. Five siting criteria were sequentially applied to a 5 km × 5 km grid, resulting in 194,027 technically feasible sites. Land-use definitions are provided in Appendix Table A3. Data sources for protected areas, topographic slope, seismic risk, and proximity to load centers (represented by nighttime lights) are listed in Appendix Table A4

**2.3 The SRO model**

For each technically feasible site, the SRO model evaluates the investment decision for a CCUS-ready coal power project under uncertainty. Investment is treated as an irreversible decision in which investors choose between immediate investment and deferral. Immediate investment realizes the project value under current market conditions, whereas deferral preserves the flexibility to respond to future changes in carbon prices, fuel prices, operating conditions, technological progress, and policy support. The optimal investment timing is therefore determined by comparing the immediate investment value with the continuation value of waiting. Spatial heterogeneity enters the model through site-specific project cash flows, thereby influencing investment timing and option value.

### 2.3.1 Operational workflow

The SRO model evaluates each technically feasible site independently by combining stochastic simulation with an optimal-stopping procedure (Fig. 3). For each site, province-specific economic and operational conditions together with site-specific source–sink relationships are first incorporated into the project cash flows. Monte Carlo simulation is then used to generate future trajectories of the stochastic variables, while the Least Squares Monte Carlo (LSMC) method estimates the continuation value at each decision point. Comparing the continuation value with the immediate investment value determines whether the project should be undertaken or deferred. Repeating this procedure across all simulated paths and technically feasible sites produces spatially explicit estimates of optimal investment timing, investment probability, and expected project value.

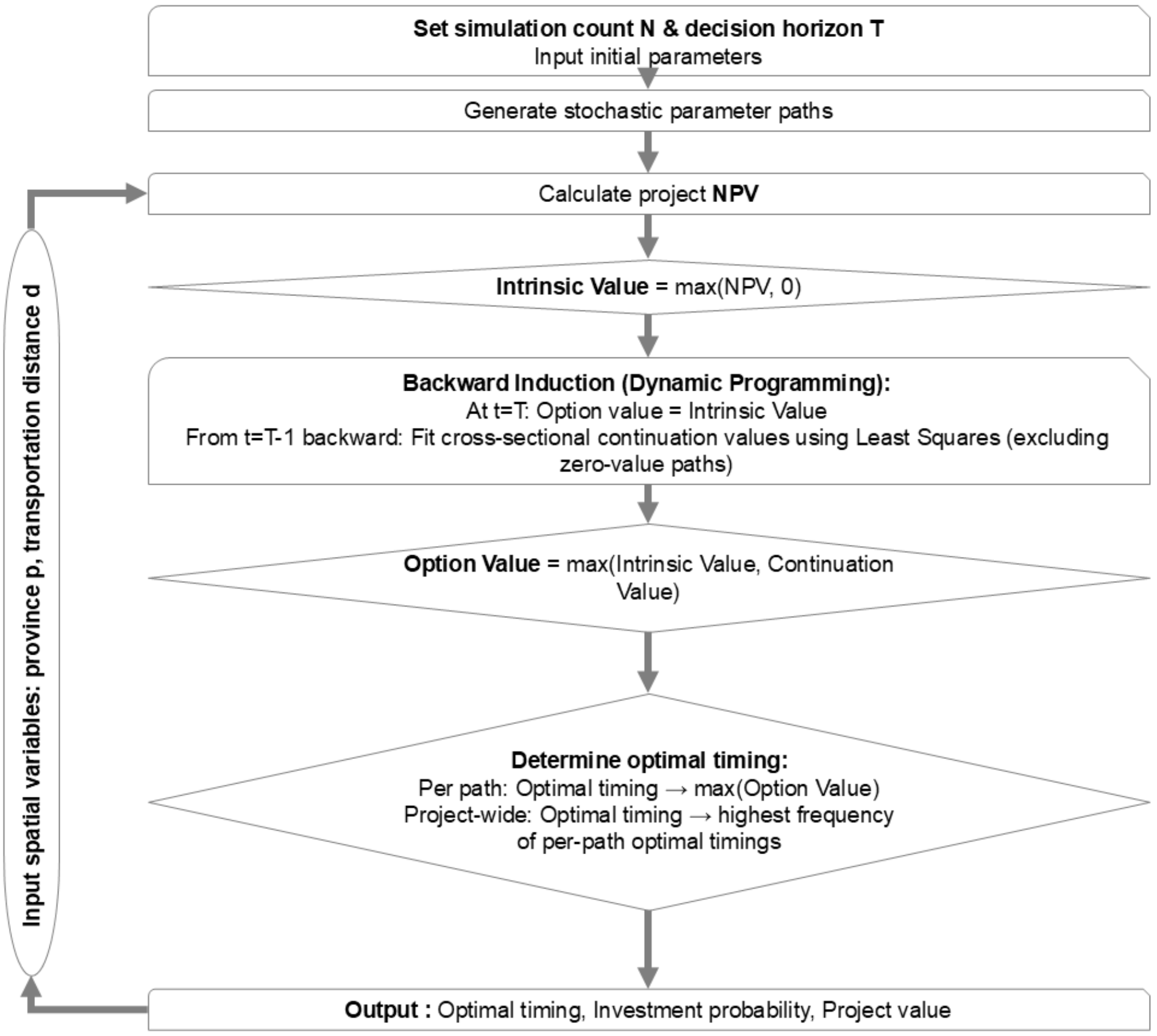

Fig. 3 Operational workflow of the Spatial Real Options (SRO) model. Site-specific economic conditions and source–sink relationships are first incorporated into project cash flows. Monte Carlo simulation and Least Squares Monte Carlo (LSMC) are then used to estimate continuation values and determine the optimal investment timing under uncertainty. Repeating

the procedure for all technically feasible sites produces national-scale maps of optimal investment timing, investment probability, and expected project value.

**2.3.2 Cash flow and valuation**

Project cash flows are simulated over a 35-year operating lifetime, with investment decisions allowed between 2025 and 2060. For each investment year, project cash flows are evaluated under the corresponding market, technology, and policy conditions. Annual project cash flow is represented as the sum of the coal power system and the CCUS system (Eq. 1). Total revenues include electricity sales, carbon trading, carbon storage revenues, and electricity subsidies for CCUS projects, whereas total costs include coal fuel and operation and maintenance (O&M) costs, CCUS O&M costs, additional fuel and electricity consumption caused by the energy penalty, and $CO_2$ transportation and storage costs. Four $CO_2$ storage options are considered: enhanced oil recovery (EOR), enhanced gas recovery (EGR), enhanced coalbed methane (ECBM), and enhanced water recovery (EWR), corresponding to oil-reservoir, gas-reservoir, coal-seam, and saline-aquifer storage, respectively. Stochastic uncertainty is represented through carbon prices, coal prices, annual generation hours, and oil or gas prices where applicable, while operating parameters and policy settings are treated as deterministic inputs. For conventional coal projects without CCUS integration, carbon costs are deducted from annual cash flows when $CO_2$ emissions exceed the free allowance, with excess emissions valued at the prevailing carbon price. For CCUS projects, captured $CO_2$ instead generates carbon-trading revenue within the CCUS system. The coal plant is valued over its 35-year operating lifetime, whereas the CCUS retrofit is valued over a 20-year operating lifetime to determine its optimal investment timing.

At time point t, the annual cash flows of the coal power system and the CCUS system are calculated as:

$$CF_t = \begin{cases} CF_t^C = R_t^E - C_t^{C,OM} - C_t^F & C \\ CF_t^{CCUS} = R_t^{CM} + R_t^S + R_t^{SE} - C_t^{CCUS,OM} - C_t^{\Delta F} - C_t^{\Delta E} - C_t^{CT} - C_t^{CS} & CCUS \end{cases} \tag{1}$$

where each revenue and cost component is calculated as:

$$\begin{cases} R_t^E = P^e \cdot Q_e \\ C_t^{C,OM} = C^{c,fom} + C^{c,vom} \\ C_t^F = P^f \cdot Q_f \\ R_t^{CM} = P^{cm} \cdot Q_{cm} \\ R_t^S = P^s \cdot Q_s \\ R_t^{SE} = P^e \cdot Q_{se} \\ C_t^{CCUS,OM} = C^{ccus,fom} + C^{ccus,vom} \\ C_t^{\Delta F} = P^{\Delta f} \cdot Q_{\Delta f} \\ C_t^{\Delta E} = P^{\Delta e} \cdot Q_{\Delta e} \\ C_t^{CT} = P^{CT} \cdot D \cdot Q_s \\ C_t^{CS} = P^{CS} \cdot Q_s \end{cases} \tag{2}$$

where $CF_t$ denotes the annual project cash flow, consisting of the cash flow from the coal-fired power system ($CF_t^C$) and the CCUS system ($CF_t^{CCUS}$). The coal power system includes electricity revenue, fuel costs, and fixed and variable O&M costs, whereas the CCUS system includes revenues from carbon trading, carbon storage, and electricity subsidies, together with CCUS O&M costs, additional fuel and electricity consumption caused by the energy penalty, and $CO_2$ transportation and storage costs. $CO_2$ transportation cost is determined by the transportation distance ($D$) between the emission source and the matched storage site. Other variables and parameters are provided in **Appendix Tables A1-A2**.

Conventional net present value (NPV) analysis evaluates project profitability by discounting future cash flows but assumes that investment occurs immediately, and therefore cannot capture the value of waiting under uncertainty (Wu et al., 2025). To address this limitation, stochastic drivers are incorporated into the cash-flow valuation, and the investment decision is formulated as an optimal-stopping problem solved using the Least Squares Monte Carlo (LSMC) method (Wei, 2025). At each decision point, the continuation value of waiting is compared with the value of immediate investment to determine the optimal investment timing and the corresponding option value for each technically feasible site.

### 2.3.3 Uncertainty measures and numerical solution

Uncertainty is incorporated into the SRO model through the principal drivers affecting project cash flows, including carbon price, coal price, annual generation hours, technological progress, oil price (for $CO_2$-EOR projects), and gas price (for $CO_2$-EGR projects). Carbon price, coal price, and generation hours directly influence project revenues and operating costs, whereas oil and gas prices determine the additional revenues associated with $CO_2$ utilization. Technological progress affects the future investment cost of CCUS deployment. The corresponding uncertainty representations are described below.

**(1) Carbon price**

Following prior work (Zhou et al., 2021), carbon price is modeled using a Geometric Brownian Motion (GBM) process:

$$\frac{dp_{c,t}}{p_{c,t}} = \mu dt + \sigma dz \tag{3}$$

where $\mu$ is the carbon price drift rate, i.e., the average growth rate, $\sigma$ is the volatility of the stochastic process of carbon price movement, and $dz$ is the increment of a standard Brownian motion. The parameters are estimated from historical carbon market data using Eqs. (4) – (9). Based on the estimated parameters, 10,000 carbon-price trajectories are generated (Fig. 4).

$$\mu_t = ln\left(\frac{p_{c,t}}{p_{c,t-1}}\right), (t = 0,1,2, \dots, n) \tag{4}$$

$$S = \sqrt{\frac{1}{n-1}\sum_{k=1}^{n} (\mu_k - \bar{\mu})^2} \tag{5}$$

$$\bar{\mu} = \frac{1}{n}\sum_{k=1}^{n} \mu_k \tag{6}$$

$$S = \sqrt{\frac{1}{n-1}\sum_{k=1}^{n} {\mu_k}^2 - \frac{1}{n(n-1)}(\sum_{k=1}^{n} \mu_k)^2} \tag{7}$$

$$\sigma = \frac{S}{\sqrt{\Delta t}} \tag{8}$$

$$\mu = \frac{\bar{\mu}}{\sqrt{\Delta t}} \tag{9}$$

The definitions and values of all model variables and parameters are provided in **Appendix Tables A1-A2**.

**(2) Coal price**

Although coal prices exhibit seasonal fluctuations, medium- and long-term transaction prices are bounded by policy (e.g., NDRC (Chinese National Development and Reform Commission) guidance in "Notice on Further Improving the Coal Market Price Formation Mechanism"), with short-term deviations typically within ±20%. We therefore model the rate of change in coal price with a bounded triangular distribution, whose probability density function is given in (Eq. 10).

$$f(x; c_{min}, c_{max}, c_{mid}) = \begin{cases} 0 & \text{for } x < c_{min} \\ \dfrac{2(x - c_{min})}{(c_{max} - c_{min})(c_{mid} - c_{min})} & \text{for } c_{min} \le x < c_{mid} \\ \dfrac{2(c_{max} - x)}{(c_{max} - c_{min})(c_{max} - c_{mid})} & \text{for } c_{mid} \le x \le c_{max} \\ 0 & \text{for } x > c_{max} \end{cases} \tag{10}$$

where, $c_{min}$, $c_{max}$, $c_{mid}$ denote the minimum, most likely, and maximum values, which are set to 0.8, 1.0, and 1.2, respectively. Based on this distribution, 10,000 coal-price trajectories are generated (Fig. 4).

**(3) Annual generation hours**

Coal-fired power plants are expected to experience declining utilization as the power system transitions toward higher shares of renewable energy while continuing to provide flexibility services. Annual generation hours are therefore modeled using a truncated exponential distribution:

$$f(x; \lambda) = \begin{cases} \lambda e^{-\lambda x} & x \ge 0, \\ 0 & x < 0, \end{cases} \tag{11}$$

where $x$ is the value of the random variable and $\lambda$ is the parameter of the distribution, often referred to as the rate parameter of the exponential distribution.

Assuming that the rate of decline in generation hours per year is random and follows a truncated exponential distribution, this study sets 5% of the initial generation level as the minimum level of generation hours to ensure a certain level of peaking and standby contingency for the unit. The variation of generation hours $H_t$ can be described by Equation (12):

$$H_{t+1} = \begin{cases} H_t - X_t & H_{t+1} \geq 5\% \cdot H_0 \\ 5\% \cdot H_0 & H_{t+1} < 5\% \cdot H_0 \end{cases} \tag{12}$$

where $t$=0 is the base year, $X_t$ is the amount of decline in generation hours in year $t$, which obeys the exponential distribution with parameter $a_t$, and its cumulative distribution function is expressed in Equation (13):

$$F(x; a_t) = 1 - e^{-a_t x} \tag{13}$$

To reflect the increasing rate of decline over time, the decline-rate parameter $a_t$ is specified by Equation (14):

$$a_t = 1\% \times \left(1 + \frac{t}{10}\right) \tag{14}$$

where 1% is the initial rate of decline and 10 is the adjustment factor controlling the speed at which the decline rate increases over time. Because the number of simulated trajectories is sufficiently large, the choice of this factor does not materially affect the final statistical results.

The simulated path of the final generation hours is shown in Fig. 4.

**(4) Technological progress**

According to the China Carbon Capture, Utilization and Storage Technology Development Roadmap (2019) (Yang et al., 2022), CCUS investment costs are expected to decline over time because of technological learning, economies of scale, and innovation. Since the timing and magnitude of future technological breakthroughs cannot be reliably represented by stochastic processes, technological progress is modeled using scenario-specific exogenous cost reduction pathways:

$$I_{t+1} = I_t \cdot ICRF_t \tag{15}$$

where $I_0$ is the initial investment cost of the retrofit in year 0 (base year), and $ICRF_t$ is the investment cost reduction factor.

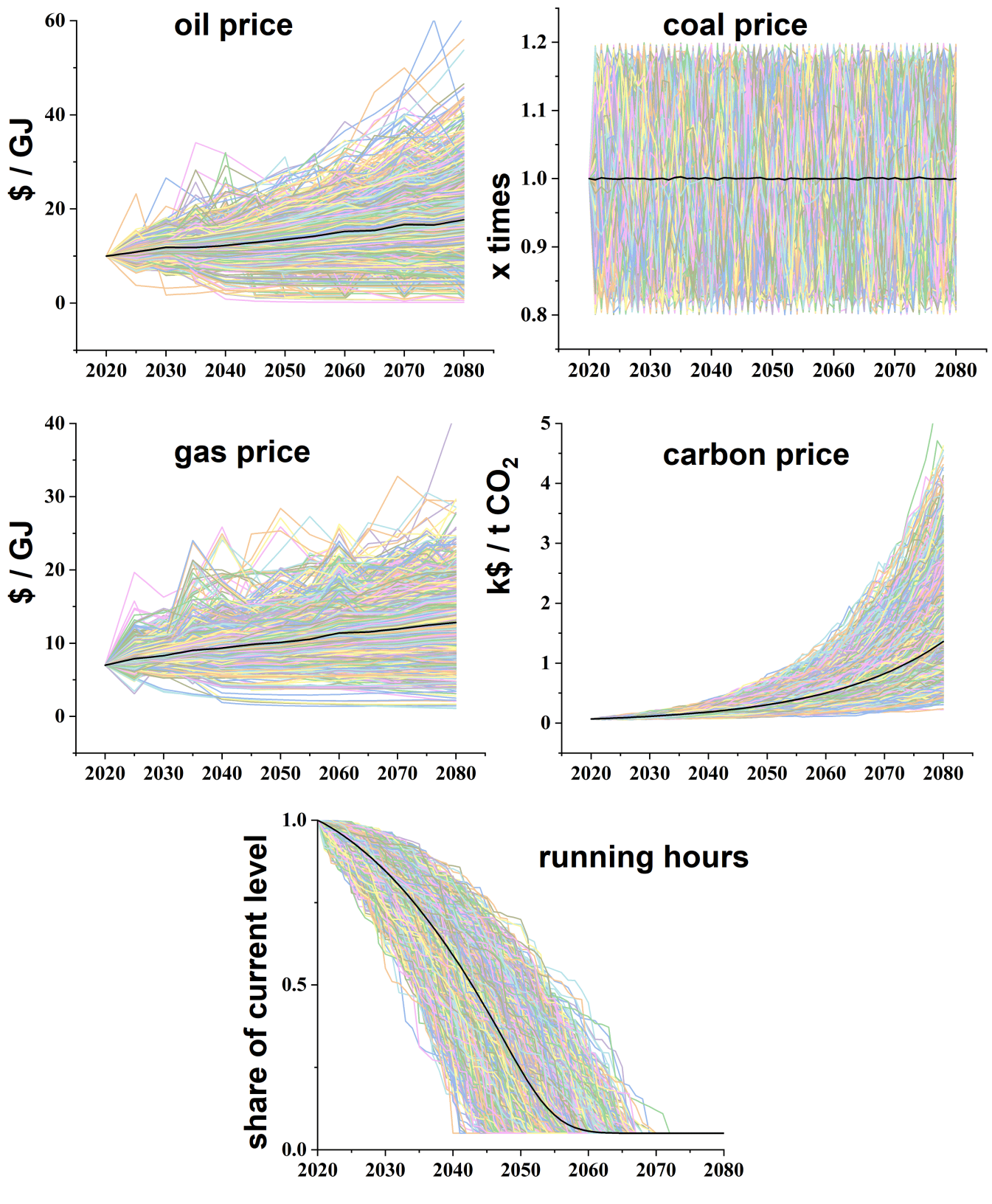

Fig. 4 Uncertainty portrayal of the key stochastic variables.

**(5) Oil price**

For $CO_2$-EOR projects, incremental revenues depend on future oil prices. Rather than assuming a specific stochastic process, this study preserves the long-term trends represented in the IPCC AR6 scenario database (Stockhause et al., 2019) while expanding the number of simulation paths through bootstrap sampling with multiplicative stochastic perturbation. Specifically, for each resampled trajectory, the perturbed oil price is generated as

$$P_t^{oil,*} = P_t^{oil}(1 + \varepsilon_t) \tag{16}$$

where $P_t^{oil}$ is the original oil price trajectory sampled from the IPCC AR6 scenario database, $P_t^{oil,*}$ is the perturbed trajectory, and $\varepsilon_t \sim N(0, \sigma_p^2)$ is a zero-mean random perturbation representing short-term deviations around the original scenario. The original 1,045 scenario trajectories are retained in the simulation set, while additional perturbed trajectories are generated to obtain 10,000 realizations for Monte Carlo simulation (Fig. 4).

**(6) Gas price**

Gas price trajectories are generated using the same procedure. Based on 1,092 natural gas price trajectories extracted from the IPCC AR6 scenario database, bootstrap resampling combined with multiplicative stochastic perturbation is applied to generate 10,000 realizations. The perturbed gas price is expressed as

$$P_t^{gas,*} = P_t^{gas}(1 + \varepsilon_t) \tag{17}$$

where the notation is analogous to Eq. (16), and $\varepsilon_t \sim N(0, \sigma_p^2)$denotes the multiplicative perturbation term.

All stochastic variables are discretized and incorporated into the Least Squares Monte Carlo (LSMC) algorithm. At each decision point, the continuation value of waiting is compared with the value of immediate investment to determine the optimal investment timing and the corresponding option value for each technically feasible site.

**2.3.4 Solving for real option value and determining optimal timing**

The optimal investment timing is determined using the Least Squares Monte Carlo (LSMC) method proposed by Longstaff and Schwartz (Longstaff and Schwartz, 2001). For each simulated path, the valuation proceeds through four steps: (i) calculating the project present value, (ii) determining the immediate exercise value, (iii) estimating the continuation value by least-squares regression, and (iv) identifying the optimal stopping time through backward induction.

Project cash flows are first converted into present values by recursively discounting future cash flows from the terminal year. The present value at each decision year is calculated as

$$PV_t = \begin{cases} CF_t, t = T \\ CF_t + PV_{t+1} \cdot e^{-r\mathrm{d}t}, t = T-1, \dots, 1 \end{cases} \tag{18}$$

where $PV_t$ is the present value at decision year $t$, $CF_t$ is the annual net cash flow obtained from Eq. (1), $r$ is the discount rate, $T$ denotes the project lifetime (35 years for conventional coal projects and 20 years for CCUS retrofit projects), and $dt$ is the annual time step.

The net present value (NPV) obtained by exercising the investment option immediately is then calculated as

$$NPV_t = PV_t - I_t, t = T, \dots, 1 \tag{19}$$

where $I_t$ denotes the investment cost at year $t$. For conventional coal projects, $I_t = I_t^c$, whereas for CCUS retrofit projects, $I_t = I_t^{cc}(1 - IR_t^{c_sub})$, where $IR_t^{c_sub}$ is the investment subsidy rate.

For Monte Carlo simulation path $i$, the intrinsic (immediate exercise) value is therefore

$$EV_{t,i} = \max\{NPV_{t,i}, 0\} \tag{20}$$

which represents the payoff obtained if the investment is undertaken immediately.

Unlike conventional discounted cash-flow analysis, real-option valuation accounts for the value of waiting under uncertainty. Following Longstaff and Schwartz (Longstaff and Schwartz, 2001), the continuation value is estimated using Least Squares Monte Carlo (LSMC). At each decision year, the discounted option values realized in the following period are regressed against polynomial basis functions of the state variables to approximate the conditional expectation of future project value.

The continuation value is approximated by

$$CV_{t,i} = \sum_{k=1}^{K} \alpha_{k,t,s} P_{t,i,s}^{k-1} + e_{t,i}, \qquad i = 1, \dots, N \tag{21}$$

where $CV_{t,i}$ denotes the continuation value for simulation path $i$ at decision year $t$, $P_{t,i,s}$ is the $s$-th state variable (carbon price, coal price, annual generation hours, oil price, or gas price), $\alpha_{k,t,s}$ are the regression coefficients, and $e_{t,i}$ is the regression residual.

The regression coefficients are estimated by ordinary least squares through minimizing the residual sum of squares,

$$min \sum_{i=1}^{N} e_{t,i}^2 \tag{22}$$

which corresponds to the polynomial approximation:

$$\alpha_{1,t,s} + \alpha_{2,t,s} P_{t,i,s} + \alpha_{3,t,s} P_{t,i,s}^2 + \cdots + \alpha_{k,t,s} P_{t,i,s}^{k-1} + e_{t,i} = e^{-r*dt} * V_{t+1,i}, \qquad i = 1, \dots, N \tag{23}$$

where $V_{t+1,i}$ is the option value in the following decision year. Polynomial basis orders from one to four were evaluated using out-of-sample cross-validation together with $R^2$ and information criteria (Brandimarte, 2013). A quadratic polynomial ($K = 2$) achieved the best balance between predictive accuracy and model parsimony and was therefore adopted.

Once the continuation value has been estimated, the option value is determined recursively by comparing the intrinsic value with the continuation value,

$$V_{t,i} = \begin{cases} EV_{t,i}, & t = T, \\ max\, EV_{t,i}, e^{-r*dt} * E\left[V_{t+1,i} \middle| p_{t,i,1}, p_{t,i,2}, \dots, p_{t,i,s}\right], & t = T-1, \dots, 1, \end{cases} \tag{24}$$

where the second term represents the continuation value of postponing investment.

Backward induction proceeds recursively from the terminal year to the initial decision year. For each simulation path, the investment is exercised when the intrinsic value first exceeds the continuation value, and the corresponding decision year is recorded as the optimal stopping time $t_i^*$. The realized option value is therefore

$$\bar{V}_{t_i^*,i} = max\{V_{t_i^*,i}, 0\} \tag{25}$$

Finally, the expected project option value is calculated by averaging the discounted option values across all Monte Carlo simulation paths,

$$F = \frac{1}{N} \sum_{i=1}^{N} e^{-rt_i^*} \bar{V}_{t_i^*,i} \tag{26}$$

where $N$ is the total number of Monte Carlo simulation paths and $t_i^*$ denotes the optimal investment time identified for simulation path $i$. The empirical distribution of $t_i^*$ across all simulation paths is subsequently used to determine the most probable investment timing for each project.

## 3. Scenarios and data sources

### 3.1 Scenario design

To investigate how technological learning, geological storage options, and policy incentives jointly influence the investment timing of CCUS-ready coal power projects, a structured scenario framework was developed. The scenario design distinguishes technology learning pathways, storage media, and policy support mechanisms, while maintaining consistent market uncertainty assumptions across all simulations. This framework enables the independent and combined effects of technology progress, storage options, and policy interventions on project investability to be systematically evaluated.

To represent technological progress, two exogenous cost-reduction pathways are adopted based on the China CCUS Technology Development Roadmap (2021) (Zhang et al., 2022): P1 (First-generation technology) and P2 (Second-generation technology). As illustrated in Fig. 5, these trajectories describe the projected decline in $CO_2$ capture and compression costs resulting from technological innovation, learning-by-doing, and economies of scale. The technology pathways are applied consistently across all scenarios to isolate the influence of technological learning from other sources of uncertainty.

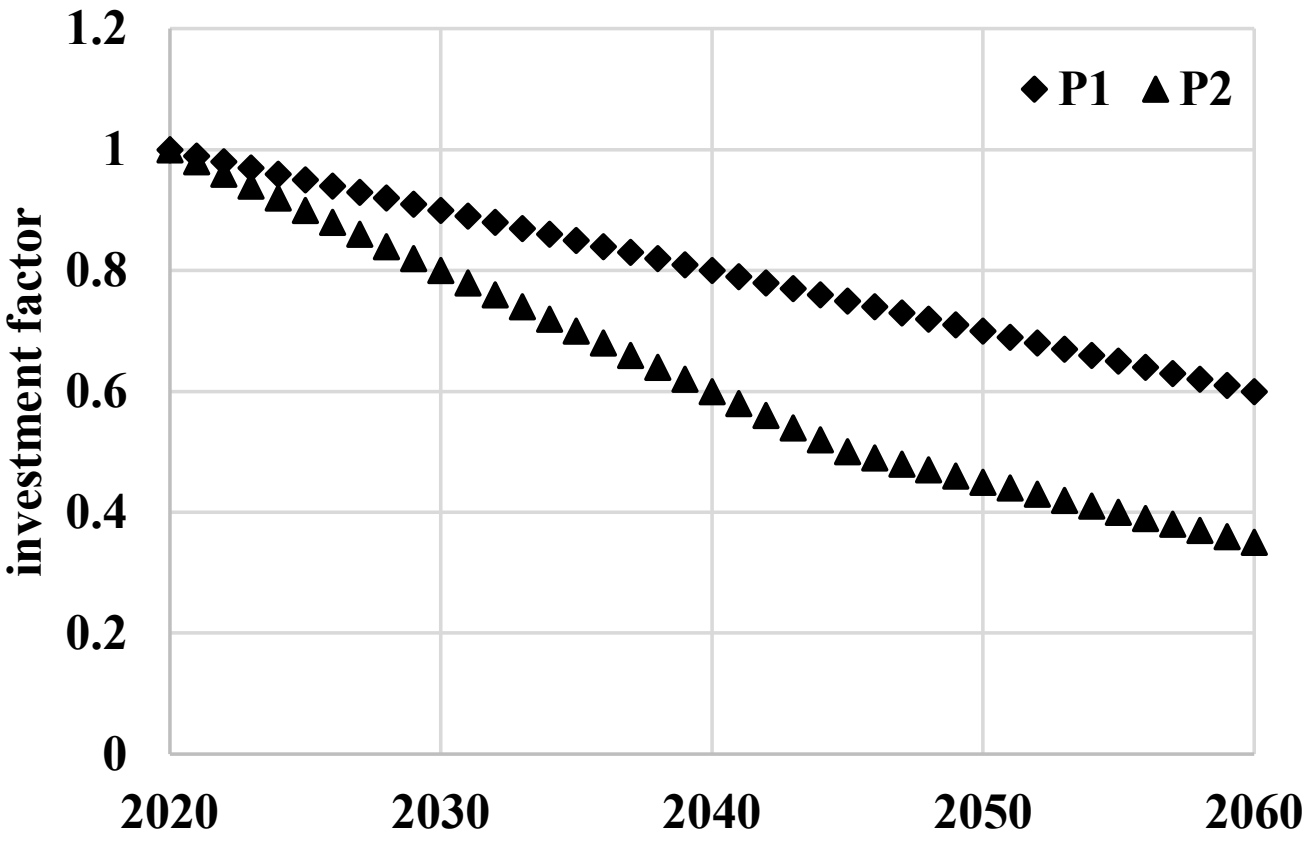

Fig. 5 Exogenous technology learning curves for CCUS capture and compression costs. Scenarios P1 (first-generation technology) and P2 (second-generation technology) represent projected cost reductions driven by technological innovation and development through 2060.

To evaluate the influence of geological storage conditions, four storage media scenarios (**S1–S4**) are considered, including EOR, EGR, ECBM, and EWR. These storage options differ substantially in storage capacity, transportation requirements, sequestration costs, and potential revenue generation, thereby providing a comprehensive basis for evaluating how storage characteristics affect project investment timing and economic performance.

Policy support is represented through two fiscal instruments that have been widely discussed in China's CCUS policy framework: generation-hour subsidies (Li et al., 2025) and investment-cost subsidies (Wei, 2025). Unlike the U.S. 45Q tax credit (Ren et al., 2022), which provides per-ton incentives but is often mutually exclusive with carbon market revenues, this study explicitly treats carbon pricing as an endogenous market uncertainty within the real-options framework. Consequently, the policy scenarios focus exclusively on subsidy mechanisms, while carbon-price uncertainty is incorporated through stochastic simulation.

To further investigate the effectiveness of alternative policy designs, Policy Sensitivity Scenarios (SW1–SW8) are established based on saline aquifer storage, which represents the largest long-term geological storage resource in China. These scenarios combine the two technology learning pathways (P1 and P2) with the two subsidy mechanisms to evaluate how technological progress and policy incentives jointly influence optimal investment timing and project value. The complete scenario definitions are summarized in Table 1.

In addition, for conventional coal projects without CCUS integration, two carbon-constraint settings are evaluated: a baseline setting in which the free carbon allowance fully covers annual emissions, and a stringent setting in which annual emissions exceed the free allowance by 20%, requiring the purchase of additional allowances at the prevailing carbon price.

Table 1 Scenario Settings

| Storage | Scenario No. | Investment cost reduction | Generation Subsidy | Investment Subsidy |
|---|---|---|---|---|
| EOR, EGR, ECBM, and EWR | S1 | P1 | 20% | 0 |
| | S2 | P2 | 20% | 0 |
| | S3 | P1 | 0 | 0 |
| | S4 | P2 | 0 | 0 |
| EWR | SW1 | P1 | 20% | 20% |
| | SW2 | P2 | 20% | 20% |
| | SW3 | P1 | 20% | 50% |
| | SW4 | P2 | 20% | 50% |
| | SW5 | P1 | 0 | 20% |
| | SW6 | P2 | 0 | 20% |
| | SW7 | P1 | 0 | 50% |
| | SW8 | P2 | 0 | 50% |

Note: For S1-S4, the following 16 scenario abbreviations are indicated to distinguish the four types of sequestration sites: S1_oil, S1_gas, S1_coal, S1_water, S2_oil, S2_gas, S2_coal, S2_water, S3_oil, S3_gas, S3_coal, S3_water, S4_oil, S4_gas, S4_coal, S4_water.

**3.2 Data sources and preprocessing**

The Spatial Real Options (SRO) framework integrates province-specific economic data with high-resolution geospatial datasets to represent both spatial heterogeneity and temporal uncertainty in CCUS-ready coal power investment. All geospatial datasets and the non-carbon economic parameters were harmonized to a common baseline around 2020 and projected onto a 5 km × 5 km spatial grid. This baseline was selected to ensure consistency among the geospatial layers while avoiding abnormal market fluctuations associated with the COVID-19 pandemic. The initial carbon market trading price, by contrast, is set to the 2024 annual average (68.15 yuan/t, Table A1) to reflect the most recent observed market level as the starting point of the stochastic carbon-price process.

Provincial coal prices were represented by official provincial coal price indices. To reduce the influence of short-term market volatility, five-year average prices during **2015–2019** were adopted. On-grid electricity prices were obtained from the official benchmark tariffs for coal-fired electricity generation in each province.

Initial operating conditions were characterized using provincial average annual generation hours during **2015–2020**, reflecting regional differences in electricity demand, power system utilization, and dispatch patterns across China's provincial grids. These values provide the baseline operating conditions for projecting future generation under uncertain market environments.

Spatial information was derived from multiple national geospatial datasets, including land-use constraints, terrain conditions, seismic risk, and $CO_2$ source–sink matching information (Wei et al., 2022b). All spatial layers were processed within the ArcGIS platform and converted to a consistent **5 km × 5 km** spatial resolution to ensure compatibility with the national inventory of coal-fired power plants and geological storage resources(Fan et al., 2025).

Economic parameters, technology costs, policy assumptions, and geospatial information were integrated within the SRO framework to generate province-specific project cash flows for each simulated investment pathway. Together, these datasets provide a consistent national-scale representation of the technical, economic, and spatial conditions required to evaluate the investment timing and economic viability of CCUS-ready coal power projects under uncertainty. Detailed descriptions of model parameters, data sources, preprocessing procedures, and variable definitions are summarized in **Appendix Tables A1-A2**.

## 4. Results

### 4.1 Conventional coal investment challenges and investment windows

To evaluate the economic viability of conventional coal-fired power investment under future market uncertainty, the SRO model was first applied to conventional coal projects without CCUS integration. Figure 6 illustrates the spatiotemporal investment windows for new 1,000 MW conventional coal-fired power units under market uncertainty. Under the baseline scenario (Fig. 6A), economically viable investment opportunities are concentrated before approximately 2045, after which project values become negative across most provinces. For provinces with positive project values, the optimal investment timing is generally immediate rather than delayed.

Clear spatial heterogeneity is observed. Provinces including Yunnan, Gansu, Qinghai, and Sichuan remain economically non-viable throughout the study period, whereas Hainan, Inner Mongolia, Guangdong, Hebei, Shaanxi, and Shanxi retain positive project values within the investment window. Under the stringent carbon-constraint scenario, where emissions exceed free carbon allowances by 20%, the investment window contracts further (Fig. 6B). The nationwide cutoff advances from approximately 2045 to 2040, and additional provinces, including Guangxi, Ningxia, Chongqing, Henan, and Jilin, become economically non-viable.

Overall, the results indicate a substantial reduction in the investment window for conventional coal-fired power projects under increasingly stringent carbon constraints.

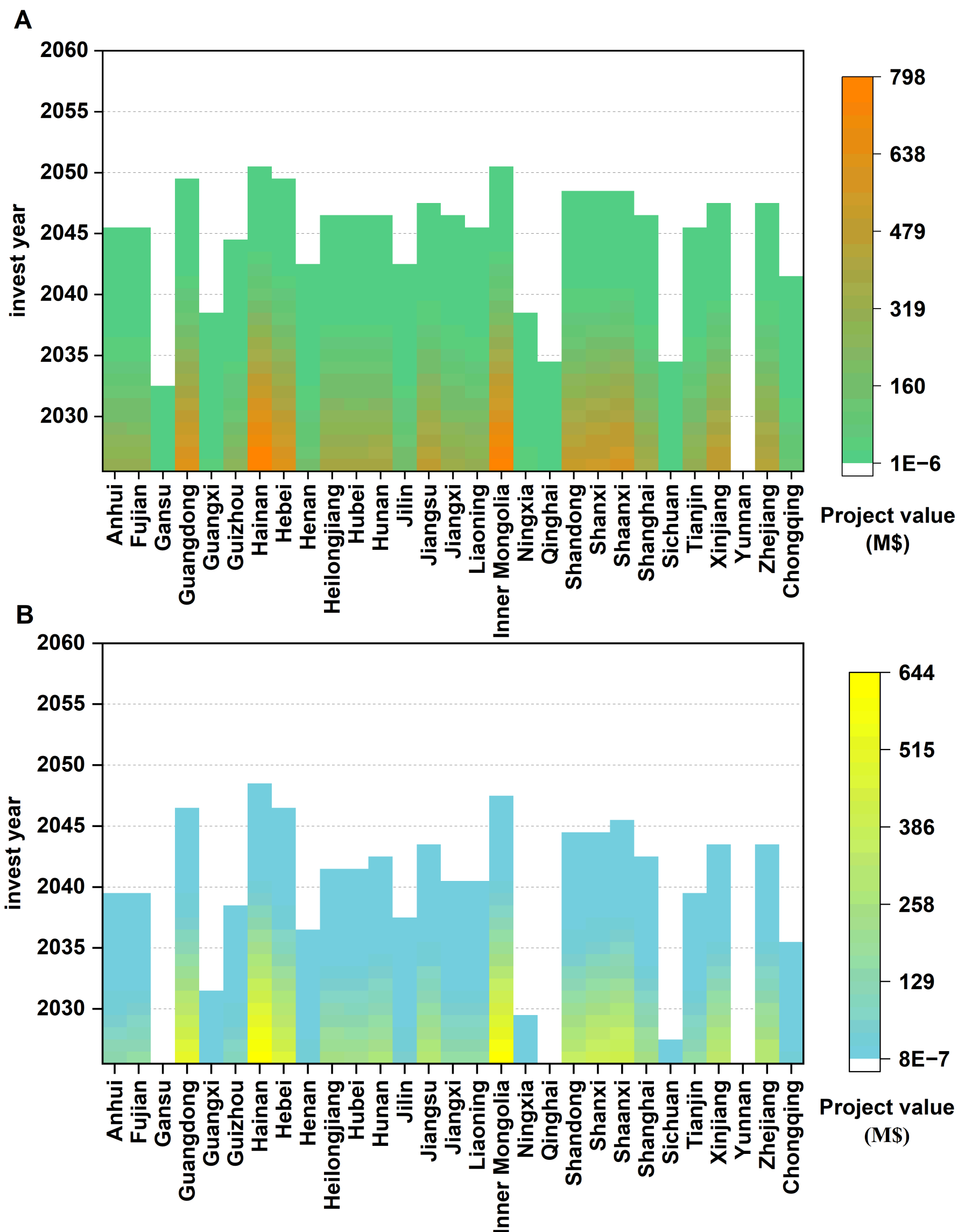

Fig. 6 Spatiotemporal investment windows and optimal investment timing for conventional coal-fired power units. (A) Baseline scenario under market uncertainty. (B) Stringent scenario

with a 20% carbon allowance shortfall. Colors indicate the optimal investment year, and gray regions denote economically non-viable projects.

### 4.2 CCUS investment decision rules

Anhui Province was selected as the representative case because it has the largest planned new coal-fired capacity among Chinese provinces, providing the most extensive basis for illustrating the investment decision rules across storage options and transport distances (Fig. 7). Figure 7A shows that the optimal investment timing varies substantially with storage type and $CO_2$ transport distance. For $CO_2$-EOR and $CO_2$-EGR, immediate investment is generally optimal when transport distances remain below 300 km. In contrast, saline aquifer storage exhibits substantially later investment timing, with optimal investment occurring after 2036 under generation-hour compensation and after 2043 without additional incentives. Coal-seam storage shows stronger sensitivity to transport distance and policy conditions. Under scenarios S1 and S2, immediate investment remains feasible when transport distances are within 150 km, whereas under S3 and S4, the optimal investment timing shifts beyond 2035 once transport distances exceed 50 km.

Figure 7B presents the expected project value evaluated at the corresponding optimal investment timing. Among all storage options, $CO_2$-EOR consistently achieves the highest project value, reaching USD 5.54 billion, followed by $CO_2$-EGR, while saline aquifer storage exhibits the lowest economic returns. The differences among storage options become more pronounced as transport distance increases. Only limited differences are observed between S1 and S2, as well as between S3 and S4. Figure 7C presents the probability of achieving a profitable investment under the simulated uncertainty. For all storage options, investment probability decreases as transport distance increases. $CO_2$-EOR maintains profitability probabilities above 80% within 100 km, whereas $CO_2$-EGR reaches similar probability levels only within 50 km under generation-hour compensation. Coal-seam storage exhibits substantially lower profitability probabilities, while saline aquifer storage remains below 7% across all scenarios. Figure 7D illustrates the temporal evolution of investment probability for the $CO_2$-EOR pathway under different transport distances. At short transport distances, high investment probabilities are concentrated in the early years. As transport distance increases, the periods with the highest investment probability gradually shift toward later years, indicating progressively delayed optimal investment opportunities.

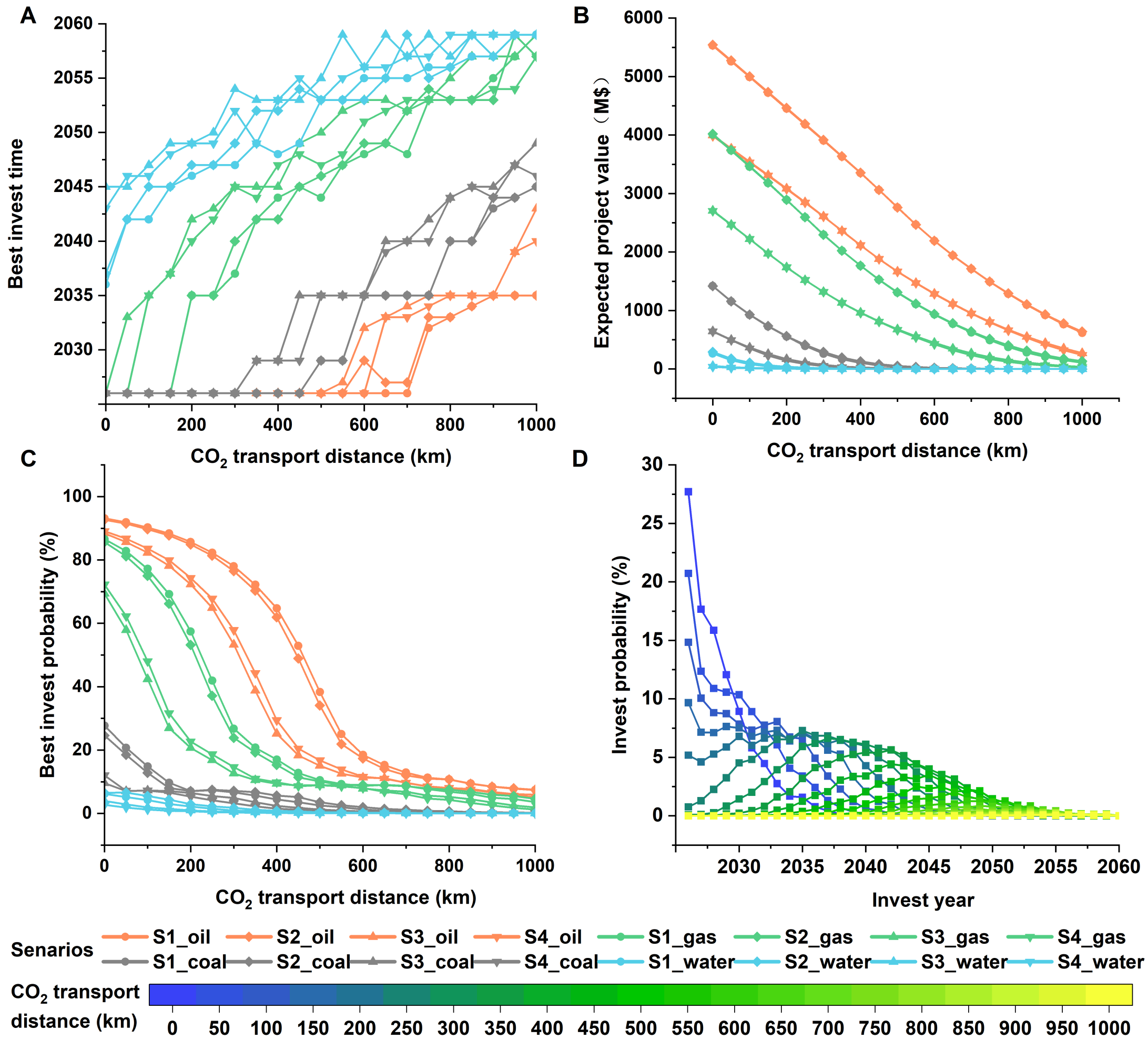

Fig. 7 Investment decision rules for CCUS-ready coal power projects in Anhui Province. (A) Optimal investment timing under different storage options and $CO_2$ transport distances. (B) Expected project value at the corresponding optimal investment timing. (C) Probability of profitable investment. (D) Temporal evolution of investment probability for the $CO_2$-EOR pathway under different transport distances. Scenario definitions: S1, normal capture-cost decline with 20% generation-hour compensation; S2, rapid capture-cost decline with compensation; S3, normal capture-cost decline without additional incentives; S4, rapid capture-cost decline without additional incentives.

4.3 National spatiotemporal mapping of CCUS investability

To investigate the national spatiotemporal distribution of CCUS-ready coal power investment, the proposed SRO framework was applied to all technically feasible sites on the 5 km × 5 km national grid under the baseline no-incentive scenario (S3). Figure 8 presents the optimal investment timing for the four storage options across China. Distinct temporal patterns are observed among different storage media. Within technically suitable areas, 90.7% and 74.4% of oil- and gas-reservoir storage sites, respectively, are optimally invested before 2030. Coal-seam storage exhibits a broader investment window, with 21.5% of sites invested before 2030 and 34.9% during 2030–2035. In contrast, saline aquifer storage is characterized by substantially later investment timing, with 54.2% of investable sites concentrated during 2035–2045. Figures 9 and 10 further summarize the investment probability and expected project value for the corresponding storage options. By 2045, the average probability of achieving a profitable investment reaches 99.5% for oil-reservoir storage and 95.9% for gas-reservoir storage, compared with 84.2% for coal-seam storage and 42.1% for saline aquifer storage. The corresponding average project values are USD 2.75, 1.40, 0.35, and 0.06 billion, respectively.

Clear spatial heterogeneity is observed across China. For all four storage options, regions with later optimal investment timing are primarily distributed along the southern coastal provinces, including Guangxi and Guangdong. In addition, the option-value distribution indicates that provinces such as Ningxia and Henan maintain positive project values under selected CCUS pathways despite relatively limited economic performance for conventional coal-fired power generation.

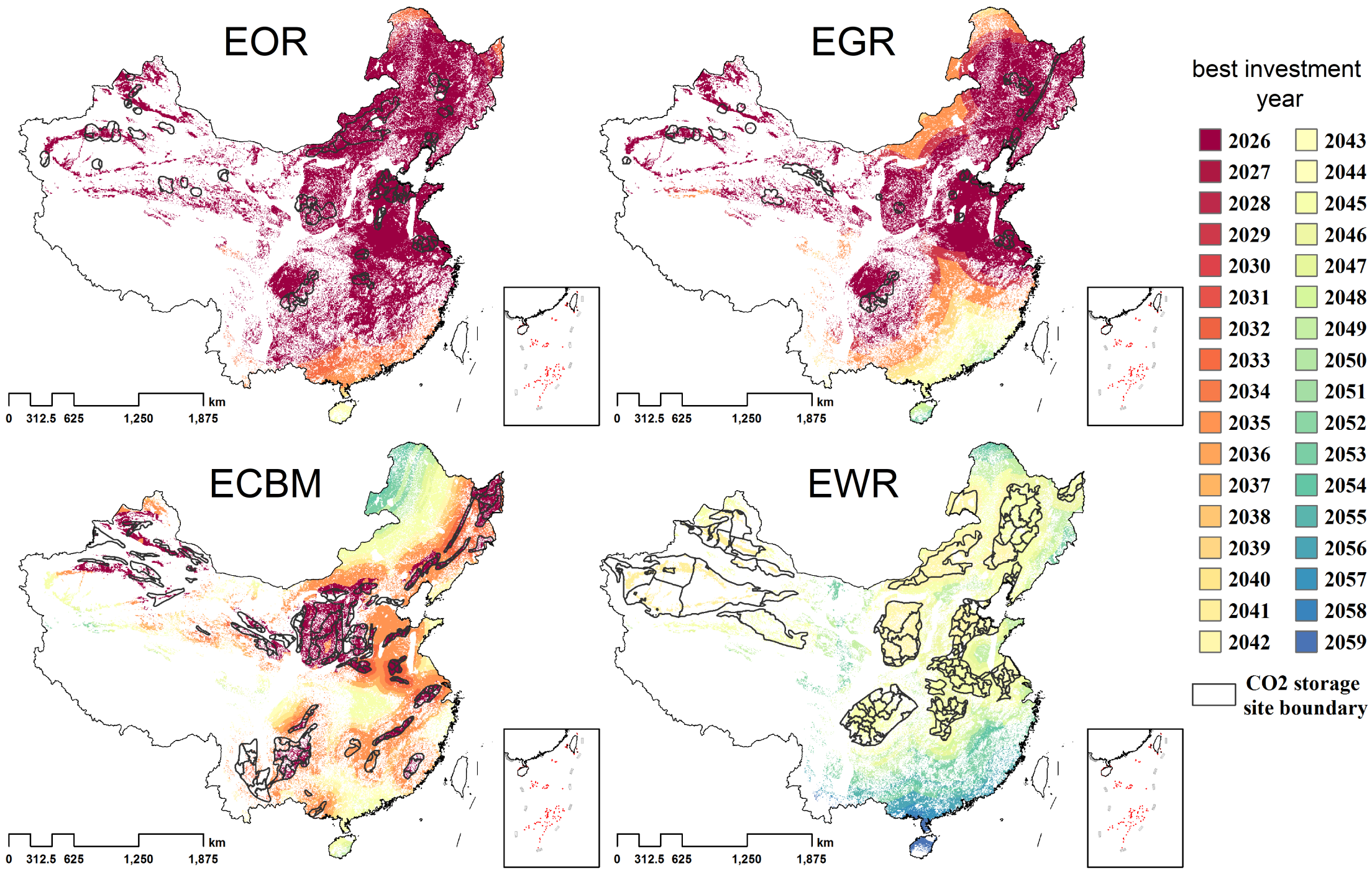

Fig. 8 National spatiotemporal distribution of optimal investment timing for CCUS-ready coal power projects under Scenario S3. Compare to subsidy scenarios, it preserves the full spatial and inter-storage heterogeneity of investment decisions. Maps show the optimal investment windows for oil-reservoir, gas-reservoir, coal-seam, and saline aquifer storage across technically suitable locations.

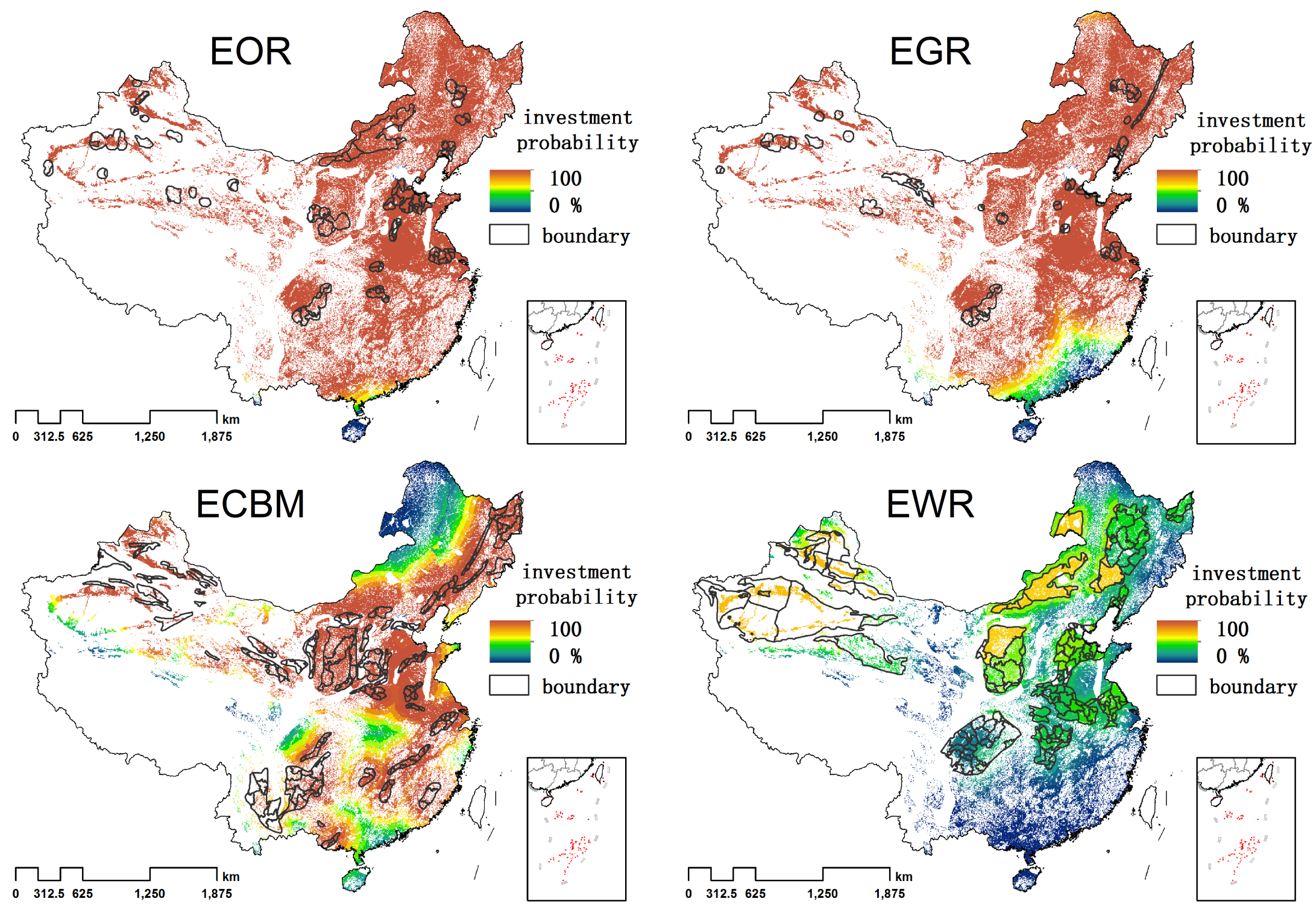

Fig. 9 Spatial distribution of the probability of profitable investment for CCUS-ready coal power projects under Scenario S3. Results represent the probability of achieving positive project value by 2045 for different storage options.

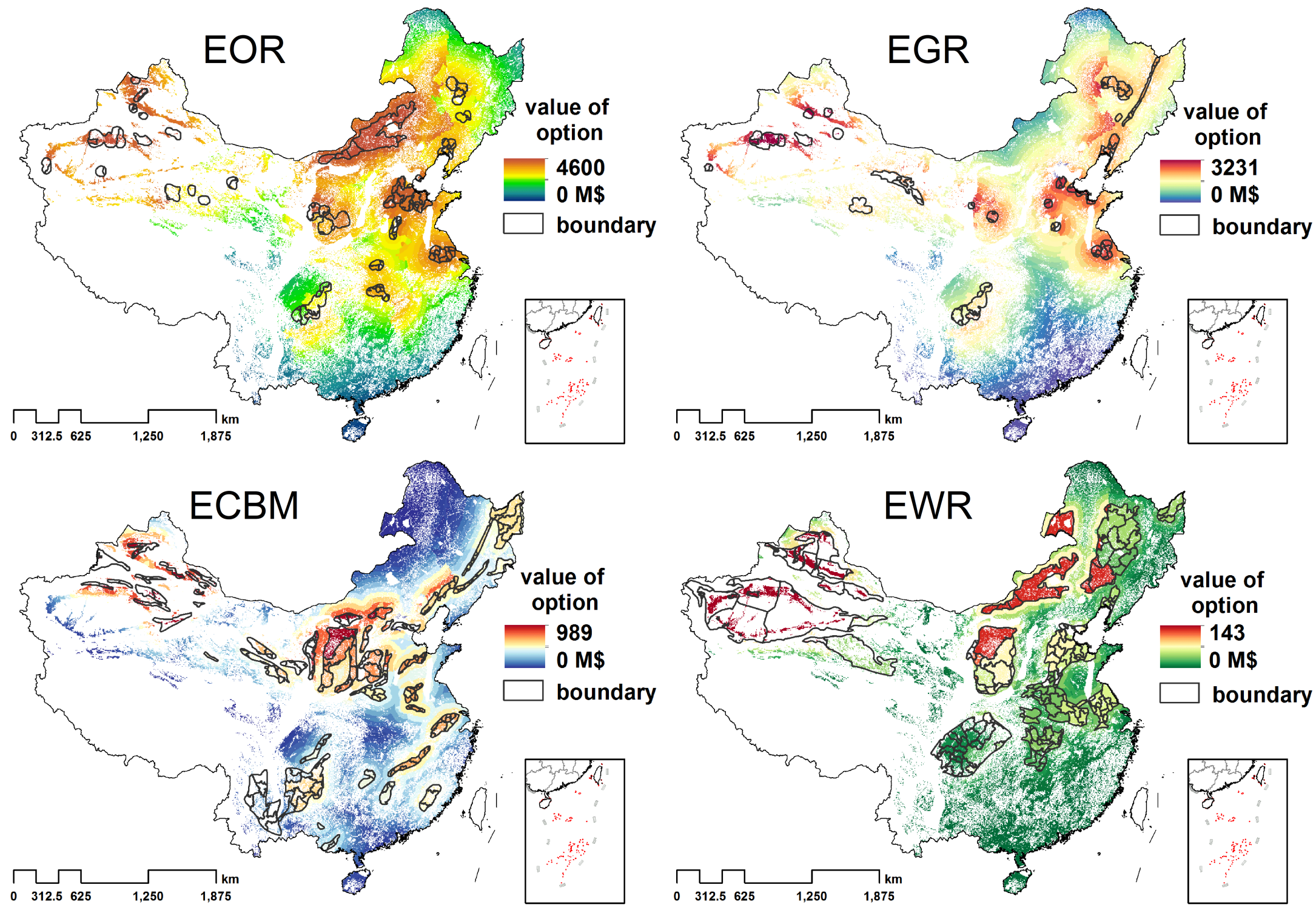

Fig. 10 Spatial distribution of the expected project value for CCUS-ready coal power projects under Scenario S3. Maps present the average project value evaluated at the corresponding

optimal investment timing.

## 4.4 Policy efficacy and incentive design

To evaluate the effectiveness of alternative policy instruments in promoting CCUS-ready coal power investment, generation-hour compensation and investment-cost subsidies were compared using saline aquifer storage as the representative pathway. Figure 11 compares the impacts of different policy instruments on investment probability and project value relative to the no-incentive baseline (S3). Generation-hour compensation consistently produces larger improvements than investment-cost subsidies across all provinces and transport distances. With a 20% generation-hour compensation, the provincial-average investment probability increases by 52% and 30% at $CO_2$ transport distances of 0 km and 100 km, respectively, while the corresponding average project value increases by USD 190 million and 60 million (Figs. 11A and 11D).

In comparison, a 20% investment-cost subsidy produces substantially smaller improvements. At transport distances of 0 km and 100 km, the average investment probability increases by 6.8% and 3.8%, respectively, while the average project value increases by USD 10 million and USD 4 million (Figs. 11B and 11E). When generation-hour compensation and investment-cost subsidies are implemented simultaneously, both investment probability and project value increase further compared with either policy instrument alone (Figs. 11C and 11F). Figure 12 further compares investment-cost subsidy rates of 20% and 50%. Increasing the subsidy rate provides only limited additional improvements in both investment probability and project value across the evaluated transport distances.

Overall, the results indicate that generation-hour compensation consistently achieves greater improvements in project investability than investment-cost subsidies under the evaluated scenarios.

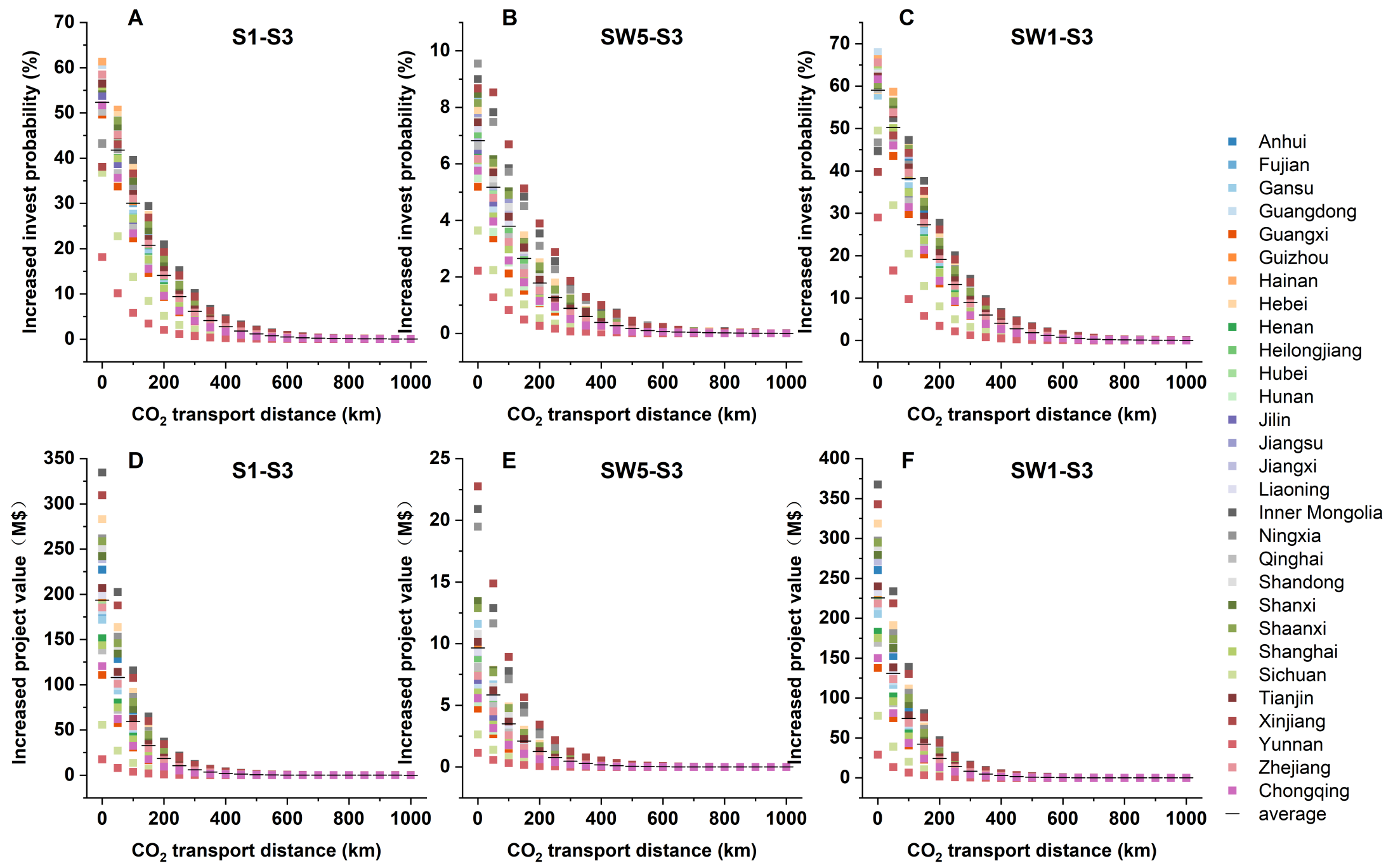

Fig. 11 Comparison of policy instruments for saline aquifer storage under different transport distances. Panels (A–C) show changes in investment probability relative to the no-incentive baseline (S3), while Panels (D–F) present the corresponding changes in project value. Positive values indicate improvements relative to Scenario S3. S1–S3 compares generation-hour compensation with the baseline, whereas SW5–S3 compares investment-cost subsidies with the baseline.

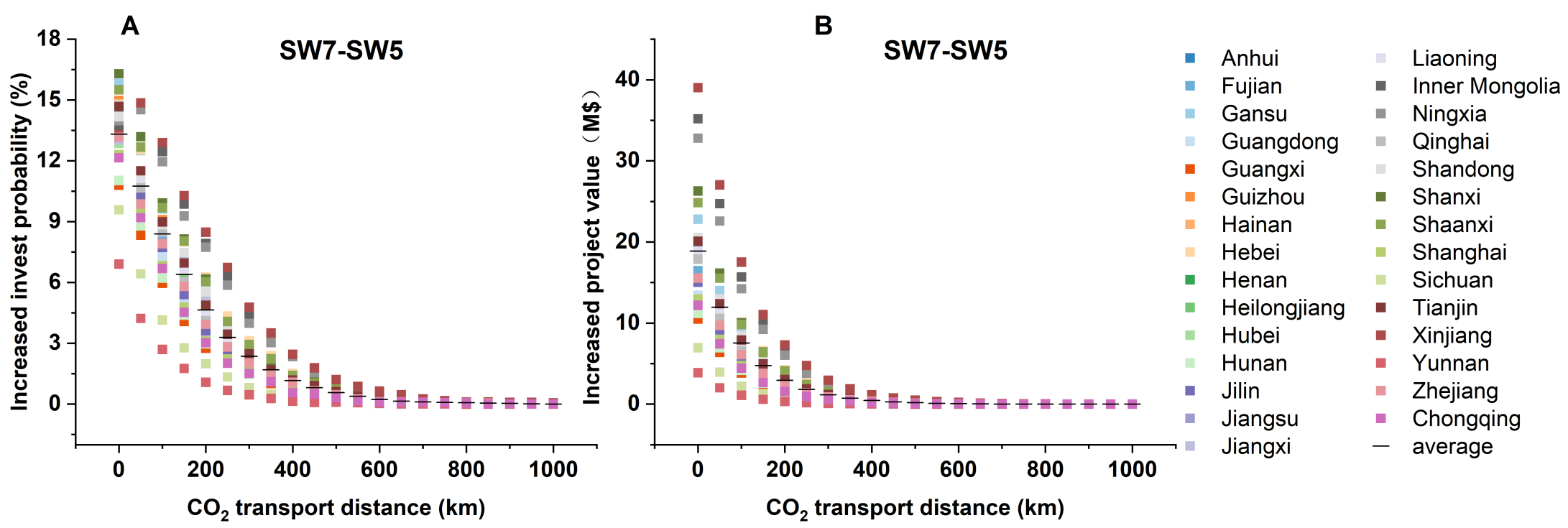

Fig. 12 Sensitivity of saline aquifer investment performance to higher investment-cost subsidy rates. Comparison between 20% (SW5) and 50% (SW7) investment-cost subsidies. (A) Incremental changes in investment probability. (B) Incremental changes in project value.

## 5. Conclusions and policy implications

This study developed a Spatial Real Options (SRO) framework to evaluate the investment timing and economic viability of CCUS-ready coal power projects under spatial heterogeneity and multiple sources of uncertainty. By integrating spatial siting constraints, technology learning, policy conditions, and stochastic market variables within a unified decision-making framework, the proposed approach provides a quantitative basis for evaluating where and when CCUS-ready coal projects should be deployed under China's carbon-neutrality pathway.

The results demonstrate that the investment window for conventional coal-fired power projects contracts rapidly and is projected to close across most regions by the early-to-mid 2040s. Incorporating CCUS substantially improves project investability, although the benefits differ considerably among storage options and transport distances. Oil- and gas-reservoir storage consistently provide higher investment value and profitability than coal-seam and saline aquifer storage, while the national spatiotemporal maps reveal pronounced regional differences in optimal investment timing. Policy evaluation further shows that generation-hour compensation is consistently more effective than investment-cost subsidies in improving project investability, particularly for saline aquifer storage.

These findings provide several implications for future CCUS deployment. First, where additional coal-fired capacity remains necessary for power system reliability, CCUS-ready design should be incorporated from the initial planning stage, with project siting coordinated with nearby geological storage resources to minimize transport costs. Second, policy support should place greater emphasis on operational performance rather than solely reducing upfront investment costs, particularly for storage options with limited direct economic returns. Third, the proposed national spatiotemporal investment maps can support regional planning by identifying priority areas where CCUS deployment is both technically feasible and economically attractive under future uncertainty.

Several limitations should be acknowledged. The current framework evaluates projects independently and therefore does not consider shared $CO_2$ transport infrastructure or competition for storage resources. In addition, the analysis assumes a fixed plant capacity and simplified geological characterization. These assumptions were adopted to maintain computational tractability for a national-scale assessment but may underestimate the benefits of infrastructure sharing and economies of scale in clustered CCUS deployment. Future work could address these limitations by integrating $CO_2$ pipeline network optimization, endogenous capacity expansion, and more detailed site-specific geological characterization within the SRO framework. Such extensions would improve the representation of large-scale CCUS deployment while providing a more comprehensive assessment of regional investment opportunities.

**Acknowledgements**

This work was supported by National Natural Science Foundation of China (grant nos. 72488101, 72293605, 72474023, 52570118, 72073014, 72104025 and 72225010).

**Author contributions**

Y.-M.W., J.-N.K., and Y.-L.Z. conceived the study. Y.-L.Z., Y.-Z.J., and S.P. contributed to the data collection and processing. Y.-L.Z. designed and performed the model runs with the help of Y.-Z.J. and S.P.; Y.-L.Z. and J.-N.K. implemented the data presentation and visualization. Y.-M.W., L.-C.L., J.-N.K., and Y.-L.Z. contributed to the interpretation of the results. Y.-L.Z. prepared the first draft. Y.-M.W., L.-C.L., and J.-N.K. worked on the review and editing. All authors approved and contributed to writing the paper.

**Competing interests**

The authors declare no competing interests.

**Lead contact**

Correspondence and requests for materials should be addressed to Prof. Yi-Ming Wei and Dr. Jia-Ning Kang.

**Appendix**

Table A1 Model parameters and input assumptions

| Item | Symbol | Numerical value (unit) | source |
|---|---|---|---|
| Coal power system | | | |
| Power plant power | | 1000 (MW) | Assumed |
| Investment Cost | $I_t^c$ | 541 (US$/kW) | (IECM Team, 2019) |
| Fixed O&M Costs | $C^{c,fom}$ | 7.18 (millions of dollars) | (IECM Team, 2019) |
| Variable O&M Costs | $C^{c,vom}$ | 3.67 (US$/MWh) | (IECM Team, 2019) |
| Coal Prices | $P^f$ | Varies by province | (Chen et al., 2021) |
| Initial Carbon Market Trading Price | $P^{cm}$ | 68.15 (yuan/tonne) | Average price in 2024 |
| Feed-in tariff | $P^e$ | Different in each province | NDRC |
| CCUS system | | | |
| Oil layer replacement rate | $ER^{C-oil}$ | 0.2 (tonnes of oil/tonnes of $CO_2$) | (Wei et al., 2022b) |
| Coal seam replacement rate | $ER^{C-methane}$ | 0.065 (tonnes methane/tonnes $CO_2$) | (Wei et al., 2022b) |
| Gas seam replacement rate | $ER^{C-gas}$ | 0.18 (tonnes of natural gas/tonnes of $CO_2$) | (Wei et al., 2022b) |
| Additional subsidized electricity | $Q_{se}$ | 20% generation | Scenario |
| Investment cost subsidy rate | $IR_t^{c_sub}$ | 20%, 50% | Scenarios |
| Carbon capture system investment cost | $I_t^{cc}$ | 262.8 (US$/kW) | (IECM Team, 2019) |
| Fixed O&M costs | $C^{ccus,fom}$ | 11.3 (US$ million) | (IECM Team, 2019) |
| Variable O&M Costs | $C^{ccus,vom}$ | 2.01 (US$/MWh) | (IECM Team, 2019) |
| Capture System Heat Consumption | $Q_{\Delta f}$ | 52.52 (tonnes of standard coal/MWh) | (IECM Team, 2019) |
| Capture system power consumption | $Q_{\Delta e}$ | 11% generation (power penalty rate) | (IECM Team, 2019) |
| $CO_2$ Pipeline transportation cost | $P^{CT}$ | 0.1 (US$/(tonne*km)) | (Wei et al., 2022b) |
| $CO_2$ Sequestration Cost | $P^{CS}$ | 6 (US$/tonne) | (Wei et al., 2022b) |
| Discount rate | $r$ | 4.7% | PBC |

Note: NDRC: National Development and Reform Commission; PBC: People's Bank of China; some of the original data in the table are in U.S. dollar units, which are converted at the RMB to U.S. dollar exchange rate of 6.8589 (the 2019 average, from the National Bureau of Statistics).

Table A2. Nomenclature of variables and symbols used in the SRO model

| Variable | Description | Unit |
|---|---|---|
| $t$ | Time step | year |
| $T$ | Project lifetime | year |
| $N$ | Number of Monte Carlo simulation paths | |
| $X_t$ | State variable used in the LSMC regression | |
| $r$ | Discount rate | |
| $\mu$ | Drift rate of the stochastic process | |
| $\sigma$ | Volatility of the stochastic process | |
| $dz$ | Increment of the Wiener process | |
| $CF_t$ | Total annual cash flow in year $t$ | US$/year |
| $NPV_t$ | The net present value in year $t$ | US$/year |
| $CF_t^C$ | Annual cash flow of the coal-fired power system | US$/year |
| $CF_t^{CCUS}$ | Annual cash flow of the CCUS system | US$/year |
| $PV_t$ | Present value of the project | US$ |
| $EV_t$ | Immediate exercise value | US$ |
| $CV_t$ | Continuation value | US$ |
| $V_t$ | Real option value | US$ |
| $R_t^E$ | Revenue from electricity generation | US$/year |
| $R_t^{CM}$ | Revenue from carbon market trading | US$/year |
| $R_t^S$ | Revenue from $CO_2$ storage or utilization | US$/year |
| $R_t^{SE}$ | Revenue from generation-hour subsidies | US$/year |
| $C_t^{C,OM}$ | Operation and maintenance cost of the coal-fired power system | US$/year |
| $C_t^F$ | Fuel cost | US$/year |
| $C_t^{CCUS,OM}$ | Operation and maintenance cost of the CCUS system | US$/year |
| $C_t^{\Delta F}$ | Additional fuel cost due to the CCUS heat penalty | US$/year |
| $C_t^{\Delta E}$ | Additional electricity cost due to the CCUS power penalty | US$/year |
| $C_t^{CT}$ | $CO_2$ transportation cost | US$/year |
| $C_t^{CS}$ | $CO_2$ storage cost | US$/year |

| Variable | Description | Unit |
|---|---|---|
| $Q_e$ | Annual electricity generation | MWh/year |
| $Q_{CO_2}$ | Annual $CO_2$ emissions | t $CO_2$/year |
| $Q_{cc}$ | Annual captured $CO_2$ | t $CO_2$/year |
| $Q_f$ | Annual coal consumption | t/year |
| $Q_{cm}$ | Verified $CO_2$ emission reduction eligible for carbon trading | t $CO_2$/year |
| $Q_S$ | Annual $CO_2$ stored | t $CO_2$/year |
| $Q_{se}$ | Electricity eligible for generation-hour subsidies | MWh/year |
| $Q_{\Delta f}$ | Additional fuel consumption due to the CCUS heat penalty | t/year |
| $Q_{\Delta e}$ | Additional electricity consumption due to the CCUS power penalty | MWh/year |
| $\eta_{cc}$ | $CO_2$ capture rate | % |
| $D$ | $CO_2$ transportation distance | km |
| $P^s$ | the revenue per unit of carbon storage, which is calculated according to the replacement rate $ER^{(C-X)}$and the price of oil and gas | US$/ t $CO_2$ |
| $ER^{(C-X)}$ | $CO_2$ utilization factor for storage option $X$(EOR, EGR or ECBM) | t product/t $CO_2$ |

Table A3 LUCC Classification System

| No. | Code | Name | Definition |
|---|---|---|---|
| 1 | Cultivated Land | | Refers to land used for growing crops, including cultivated land, newly reclaimed land, fallow land, rotational fallow land, and grassland-crop rotation land; agricultural land primarily used for growing crops, such as orchards, mulberry plantations, and agroforestry land; as well as tidal flats and coastal mudflats that have been cultivated for more than three years. |
| - | 11 | Paddy Fields | Refers to arable land with a guaranteed water supply and irrigation facilities that can be normally irrigated under typical weather conditions, used for growing rice, lotus roots, and other aquatic crops, including land where rice and dryland crops are rotated.<br>111 Mountainous paddy fields 112 Hilly paddy fields 113 Plain paddy fields 114 Paddy fields on slopes steeper than 25 degrees |
| - | 12 | Dryland | Refers to arable land without irrigation water sources or facilities, where crops are grown relying on natural rainfall; arable land for dryland crops that has water sources and irrigation facilities and can be normally irrigated under normal conditions; arable land primarily used for vegetable cultivation; and fallow land and rotational fallow land under normal crop rotation.<br>121 Mountainous dryland 122 Hilly dryland 123 Plain dryland 124 Dryland on slopes steeper than 25 degrees |
| 2 | Forest Land | | Refers to forest land where trees, shrubs, bamboo, and coastal mangrove forests grow. |
| - | 21 | Forested Land | Refers to natural and planted forests with a canopy cover of >30%. This includes contiguous forest areas such as timber forests, commercial forests, and shelterbelts. |
| - | 22 | Shrubland | Refers to low-growing forest and shrubland with a canopy cover of >40% and a height of 2 meters or less. |
| - | 23 | Open Woodland | Refers to forested areas with a canopy cover of 10–30%. |

| - | 24 | Other Forest Lands | Refers to unestablished afforestation sites, clear-cut areas, nurseries, and various orchards (fruit orchards, mulberry orchards, tea plantations, tropical crop plantations, etc.). |
|---|---|---|---|
| 3 | Grassland | | Refers to all types of grasslands dominated by herbaceous plants with a ground cover of 5% or more, including shrubby grasslands used primarily for grazing and open grasslands with a canopy cover of 10% or less. |
| - | 31 | High-Coverage Grasslands | Refers to natural grasslands, improved grasslands, and hayfields with a coverage of >50%. Such grasslands generally have favorable moisture conditions and dense vegetation. |
| - | 32 | Medium-Coverage Grassland | Refers to natural and improved grasslands with a coverage of >20–50%. These grasslands generally have insufficient moisture and sparse vegetation. |
| - | 33 | Low-coverage grassland | Refers to natural grasslands with a coverage of 5－20%. Such grasslands suffer from water scarcity, have sparse vegetation, and are unsuitable for grazing. |
| 4 | Waterbodies | | Refers to natural inland water bodies and land used for water conservancy facilities. |
| - | 41 | Rivers and canals | Refers to land located below the normal water level of naturally formed or artificially excavated rivers and main watercourses. Artificial channels include their embankments. |
| - | 42 | Lakes | Refers to land located below the normal water level of a naturally formed body of standing water. |
| - | 43 | Reservoirs and Ponds | Refers to land located below the normal water level of an artificially constructed water storage area. |
| - | 44 | Permanent Glaciers and Snowfields | Land that is permanently covered by glaciers and snow. |
| - | 45 | Tidal flats | Refers to the intertidal zone between high and low tide levels along the coast. |

| - | 46 | Floodplain | Refers to land situated between the normal water level and the floodwater level of rivers and lakes. |
|---|---|---|---|
| 5 | Urban, Rural, Industrial, Mining, and Residential Land | | Refers to land used for urban and rural settlements, as well as industrial, mining, transportation, and other purposes outside these settlements. |
| - | 51 | Urban Land | Refers to land within the built-up areas of large, medium, and small cities, as well as county towns and above. |
| - | 52 | Rural Settlements | Refers to rural settlements located outside urban areas. |
| - | 53 | Other Construction Land | Refers to land used for factories, mines, large industrial zones, oil fields, salt fields, quarries, etc., as well as transportation roads, airports, and special-purpose land. |
| 6 | Unutilized Land | | Land that is currently unused, including land that is difficult to utilize. |
| - | 61 | Sandy Land | Land with a sandy surface and vegetation cover of less than 5%, including deserts but excluding deserts within water systems. |
| - | 62 | Gobi | Land where the surface consists primarily of gravel and the vegetation cover is less than 5%. |
| - | 63 | Saline-alkali land | Land characterized by the accumulation of salt and alkali on the surface, sparse vegetation, and the ability to support only highly salt- and alkali-tolerant plants. |
| - | 64 | Wetlands | Land characterized by flat, low-lying terrain with poor drainage, which remains persistently damp and is subject to seasonal or year-round waterlogging, with wetland plants growing on the surface. |
| - | 65 | Bare Land | Land with a soil surface and vegetation cover of less than 5%. |
| - | 66 | Bare Rock | Refers to land where the surface consists of rock or gravel, with a coverage area of more than 5%. |
| - | 67 | Other | Refers to other unutilized land, including alpine deserts, tundra, etc. |

Table A4. Spatial datasets used for site screening and model parameterization

| **Data Category** | **Dataset** | **Spatial/Temporal Resolution** | **Year** | **Source** | **Download URL** |
|---|---|---|---|---|---|
| Topography and geology | Digital Elevation Model (DEM) | 1 km × 1 km | 2019 | National Geomatics Center of China (NGCC) | |
| Topography and geology | Land use | 1 km × 1 km | 1980–2020 | Resource and Environment Science and Data Center, Chinese Academy of Sciences (RESDC) | www.gisrs.cn |
| Topography and geology | Nature reserves | Polygon (shapefile) | 2021 | Ministry of Ecology and Environment of China (MEE) | |
| Topography and geology | Peak Ground Acceleration (PGA) | Polygon (shapefile) | 2015 | China Earthquake Administration (CEA) | https://www.gb18306.net |
| Socio-economic data | Nighttime lights | 1 km × 1 km | 1992–2022 | National Oceanic and Atmospheric Administration (NOAA) | https://dataverse.harvard.edu/dataset.xhtml?persistentId=doi:10.7910/DVN/GIYGJU |